\documentclass{svmult}

\usepackage{makeidx}         
\usepackage{graphicx}        

\begin{document}

\title*{Ideal-Gas Like Markets: Effect of Savings}
\author{Arnab Chatterjee and Bikas K Chakrabarti}
\institute{
Theoretical Condensed Matter Physics Division and 
Centre for Applied Mathematics and Computational Science, 
Saha Institute of Nuclear Physics, 
Block-AF, Sector-I Bidhannagar, Kolkata-700064, India.
\texttt{arnab.chatterjee@saha.ac.in, bikask.chakrabarti@saha.ac.in}}
%
%
\maketitle
\begin{abstract}
We discuss the ideal gas like models of a trading market. The effect of savings
on the distribution have been thoroughly reviewed.
The market with fixed saving factors leads to a Gamma-like
distribution. In a  market with quenched random saving factors for its agents 
we show that the steady state income ($m$) distribution $P(m)$ in the model 
has a power law tail with Pareto index $\nu$ equal to unity. We also
discuss the detailed numerical results on this model. 
We analyze the distribution of mutual money
difference and also develop a master equation for the time
development of $P(m)$. Precise solutions are then obtained in some special
cases.
\end{abstract}

\section{Introduction}
\noindent
The distribution of wealth among individuals in an economy has been an
important area of research in economics, for more than a hundred
years. Pareto \cite{cc:Pareto:1897} first quantified the high-end of the income
distribution in a society and found it to follow a power-law
\begin{equation}
\label{par}
P(m) \sim m^{-(1+\nu)},
\end{equation}
where $P$ gives the normalized number of people
with income $m$, and the exponent $\nu$ is called the Pareto index.

Considerable investigations with real data during the last ten years revealed
that the tail of the income distribution indeed follows the above
mentioned behavior and the value of the Pareto index $\nu$ is generally
seen to vary between 1 and 3 \cite{cc:Oliveira:1999,cc:realdatag,cc:realdataln,cc:Sitabhra:2005}. It is also
known that typically less than $10 \%$ of the population in any country
possesses about $40 \%$ of the total wealth of that country and they follow
the above law. The rest of the low income population, in fact the majority
($90\%$ or more), follow a different distribution which is debated to be either
Gibbs \cite{cc:realdatag,cc:marjit,cc:Dragulescu:2000} or log-normal \cite{cc:realdataln}.

Much work has been done recently on models of markets, where economic (trading)
activity is analogous to some scattering process
\cite{cc:marjit,cc:Chakraborti:2000,cc:Chakraborti:2002,cc:Hayes:2002,
cc:Chakraborti:2001,cc:Chatterjee:2004,cc:Chatterjee:2003,
cc:Chakrabarti:2004,cc:othermodels,cc:Slanina:2004} as in the kinetic theory 
\cite{cc:Landau:1968} of gases or liquids.

We put our attention to models where introducing a saving propensity (or factor)
\cite{cc:Samuelson:1980} for the
agents, a wealth distribution similar to that in the real economy can be
obtained \cite{cc:Chakraborti:2000,cc:Chatterjee:2004}. 
Savings do play an important role in
determining the nature of the wealth distribution in an economy and this
has already been observed in some recent investigations \cite{cc:Willis:2004}.
Two variants of the model have been of recent interest; namely, where the agents
have the same fixed saving factor \cite{cc:Chakraborti:2000}, and where the agents
have a quenched random distribution of saving factors \cite{cc:Chatterjee:2004}.
While the former has been understood to a certain extent (see e.g,
\cite{cc:Das:2003,cc:Patriarca:2004}), and argued to resemble a gamma distribution
\cite{cc:Patriarca:2004},
attempts to analyze the latter model are still incomplete (see however
\cite{cc:Repetowicz:2004}). 
Further numerical studies \cite{cc:Ding:2003} of time correlations in
the model seem to indicate even more intriguing features of the model.
In this paper, we intend to analyze the second market model with randomly
distributed saving factor, using a master equation type approach similar
to kinetic models of condensed matter.

We have studied here numerically a gas model of a trading market. We have
considered the effect of saving propensity of the traders. The saving
propensity is assumed to have a randomness. Our observations indicate that
Gibbs and Pareto distributions fall in the same category and can appear
naturally in the century-old and well-established kinetic theory of gas
\cite{cc:Landau:1968}: Gibbs distribution for no saving and Pareto distribution for
agents with quenched random saving propensity. Our model study also indicates
the appearance of self-organized criticality \cite{cc:Bak:1997} in the simplest 
model so far, namely in the kinetic theory of gas models, when the stability 
effect of savings \cite{cc:Samuelson:1980} is incorporated.

\section{Ideal-gas like models}
\label{idealgas}

\noindent
We consider an ideal-gas model of a closed economic system
where total money $M$ and total number of agents $N$ is fixed. No production
or migration occurs and the only economic activity is confined to trading.
Each agent $i$, individual or corporate, possess money $m_i(t)$ at time $t$.
In any trading, a pair of traders $i$ and $j$ randomly exchange their money
\cite{cc:marjit,cc:Dragulescu:2000,cc:Chakraborti:2000}, such that their total money 
is (locally) conserved
and none end up with negative money ($m_i(t) \ge 0$, i.e, debt not allowed):
\begin{equation}
\label{consv}
m_i(t) + m_j(t) = m_i(t+1) + m_j(t+1);
\end{equation}

\noindent time ($t$) changes by one unit after each trading.
The steady-state ($t \rightarrow \infty$) distribution of money is Gibbs one:
\begin{equation}
\label{gibbs}
P(m)=(1/T)\exp(-m/T);T=M/N. 
\end{equation}

Hence, no matter how uniform or justified the initial distribution is, the
eventual steady state corresponds to Gibbs distribution where most of the
people have got very little money.
This follows from the conservation of money and additivity of entropy:
\begin{equation}
\label{prob}
P(m_1)P(m_2)=P(m_1+m_2).
\end{equation}

\noindent
This steady state result is quite robust and realistic too!
In fact, several variations of the trading, and of the `lattice'
(on which the agents can be put and each agent trade with its
`lattice neighbors' only), whether compact, fractal or small-world like
\cite{cc:Oliveira:1999}, leaves the distribution unchanged. Some other variations
like random sharing of an amount $2m_2$ only (not of $m_1 + m_2$)
when $m_1 > m_2$ (trading at the level of lower economic class in the trade),
lead to even drastic situation: all the money in the market drifts to one
agent and the rest become truely pauper \cite{cc:Chakraborti:2002,cc:Hayes:2002}.

\subsection{Effect of fixed or uniform savings}
\label{fixedsaving}

\noindent
In any trading, savings come naturally \cite{cc:Samuelson:1980}.
A saving propensity factor $\lambda$ is therefore introduced in the same model
\cite{cc:Chakraborti:2000} (see \cite{cc:Dragulescu:2000} for model without 
savings), where each trader
at time $t$ saves a fraction $\lambda$ of its money $m_i(t)$ and trades
randomly with the rest:
\begin{equation}
\label{delm}
m_{i}(t+1)=m_{i}(t)+\Delta m;\quad m_{j}(t+1)=m_{j}(t)-\Delta m 
\end{equation}
\noindent where
\begin{equation}
\label{eps}
\Delta m=(1-\lambda )[\epsilon \{m_{i}(t)+m_{j}(t)\}-m_{i}(t)],
\end{equation}

\noindent
$\epsilon$ being a random fraction, coming from the stochastic nature
of the trading.
\begin{figure}
\vskip .3cm
\centering
\includegraphics[height=5cm]{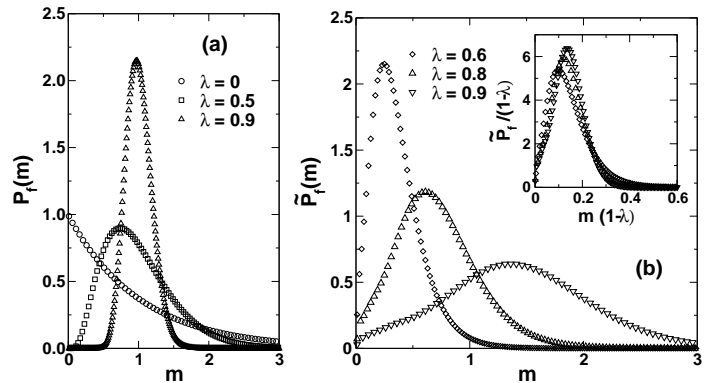}
\vskip .2cm
\caption{Steady state money distribution (a) $P(m)$ for the fixed $\lambda$
model, and (b) $\tilde{P_f}(m)$ for some specific values of $\lambda$
in the distributed $\lambda$ model. All data are for $N=200$. Inset of (b)
shows scaling behavior of $\tilde{P_f}(m)$.}
\label{ccfig:1}       
\end{figure}

The market (non-interacting at $\lambda =0$ and $1$) becomes `interacting'
for any non-vanishing $\lambda (<1)$: For fixed $\lambda$ (same for all
agents), the steady state distribution $P_f(m)$ of money is exponentially
decaying on both sides with the most-probable money per agent shifting away
from $m=0$ (for $\lambda =0$) to $M/N$ as $\lambda \rightarrow 1$
(Fig. 1(a)). This self-organizing feature of the market,
induced by sheer self-interest of saving by each agent without any global
perspective, is quite significant as the fraction of paupers decrease with
saving fraction $\lambda$ and most people end up with some fraction of the
average money in the market (for $\lambda \rightarrow 1$, the socialists'
dream is achieved with just people's self-interest of saving!).
Interestingly, self-organisation also occurs in such market models when there
is restriction in the commodity market \cite{cc:Chakraborti:2001}.
Although this fixed saving propensity does not give yet the Pareto-like
power-law distribution, the Markovian nature of the scattering or trading
processes (eqn. (\ref{prob})) is lost and the system becomes co-operative.
Indirectly through \(\lambda\), the agents get to know (start interacting with)
each other and the system co-operatively self-organises towards a most-probable
distribution (\(m_p \ne 0\)).

This has been understood to a certain extent (see e.g,
\cite{cc:Das:2003,cc:Patriarca:2004}), and argued to resemble a gamma distribution
\cite{cc:Patriarca:2004}, and partly explained analytically \cite{cc:Repetowicz:2004}.

\subsection{Effect of distributed savings}
\label{mixedsaving}

\noindent
In a real society or economy, $\lambda$ is a very inhomogeneous parameter:
the interest of saving varies from person to person.
We move a step closer to the real situation where saving factor $\lambda$ is
widely distributed within the population 
\cite{cc:Chatterjee:2004,cc:Chatterjee:2003,cc:Chakrabarti:2004}. 
\begin{figure}
\centering
\includegraphics[height=6.2cm]{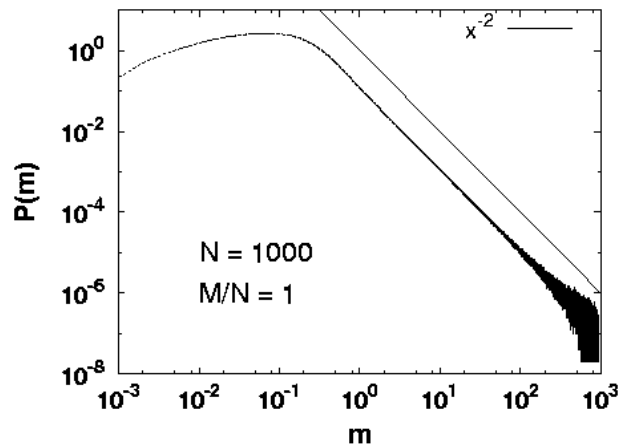}
\caption{Steady state money distribution $P(m)$ for the distributed
$\lambda$ model with $0 \le \lambda < 1$ for a system of $N=1000$ agents.
The $x^{-2}$ is a guide to the observed power-law, with $1+\nu=2$.}
\label{ccfig:2}       
\end{figure}

The evolution of money in such a trading
can be written as:
\begin{equation}
\label{mi}
m_i(t+1)=\lambda_i m_i(t) + \epsilon_{ij} \left[(1-\lambda_i)m_i(t) + (1-\lambda_j)m_j(t)\right], 
\end{equation}
\begin{equation}
\label{mj}
m_j(t+1)=\lambda_j m_j(t) + (1-\epsilon_{ij}) \left[(1-\lambda_i)m_i(t) + (1-\lambda_j)m_j(t)\right]
\end{equation}

One again follows the same trading rules as before, except that
\begin{equation}
\label{lrand}
\Delta m=\epsilon_{ij}(1-\lambda_{j})m_{j}(t)-(1-\lambda _{i})(1 - \epsilon_{ij})m_{i}(t)
\end{equation}

\noindent
here; $\lambda _{i}$ and $\lambda _{j}$ being the saving
propensities of agents $i$ and $j$. The agents have fixed (over time) saving
propensities, distributed independently, randomly and uniformly (white)
within an interval $0$ to $1$ agent $i$ saves a random fraction
$\lambda_i$ ($0 \le \lambda_i < 1$) and this $\lambda_i$ value is quenched
for each agent ($\lambda_i$ are independent of trading or $t$).
Starting with an arbitrary initial (uniform or random) distribution of
money among the agents, the market evolves with the tradings. At each time,
two agents are randomly selected and the money exchange among them occurs,
following the above mentioned scheme. We check for the steady state, by
looking at the stability of the money distribution in successive
Monte Carlo steps $t$ (we define one Monte Carlo time step as $N$ pairwise
interations). Eventually, after a typical relaxation time
($\sim 10^6$ for $N=1000$ and uniformly distributed $\lambda$) dependent
on $N$ and the distribution of $\lambda$, the money distribution becomes
stationary. After this, we average the money distribution over $\sim 10^3$
time steps. Finally we take configurational average over $\sim 10^5$
realizations of the $\lambda$ distribution to get the money distribution
$P(m)$. It is found to follow a strict power-law decay.
This decay fits to Pareto
law (\ref{par}) with $\nu = 1.01 \pm 0.02$ (Fig. 2). Note, for
finite size $N$ of the market, the distribution has a narrow initial growth
upto a most-probable value $m_p$ after which it falls off with a power-law
tail for several decades. This
Pareto law (with $\nu \simeq 1$) covers the entire range in $m$ of the
distribution $P(m)$ in the limit $N \rightarrow \infty$. We checked that
this power law is extremely robust: apart from the uniform $\lambda$
distribution used in the simulations in Fig. 2, we also checked the results
for a distribution
\begin{equation}
\label{lam0}
\rho(\lambda) \sim |\lambda_0-\lambda|^\alpha,\quad \lambda_0 \ne 1, \quad 0<\lambda<1,
\end{equation}

\noindent
of quenched $\lambda$ values among the agents. The Pareto law with $\nu=1$ is
universal for all $\alpha$. The data in Fig. 2 corresponds to
$\lambda_0 = 0$, $\alpha = 0$. For negative $\alpha$ values, however,
we get an initial (small $m$) Gibbs-like decay in $P(m)$ (see Fig. 3).
\begin{figure}
\vskip .2cm
\centering
\includegraphics[height=6cm]{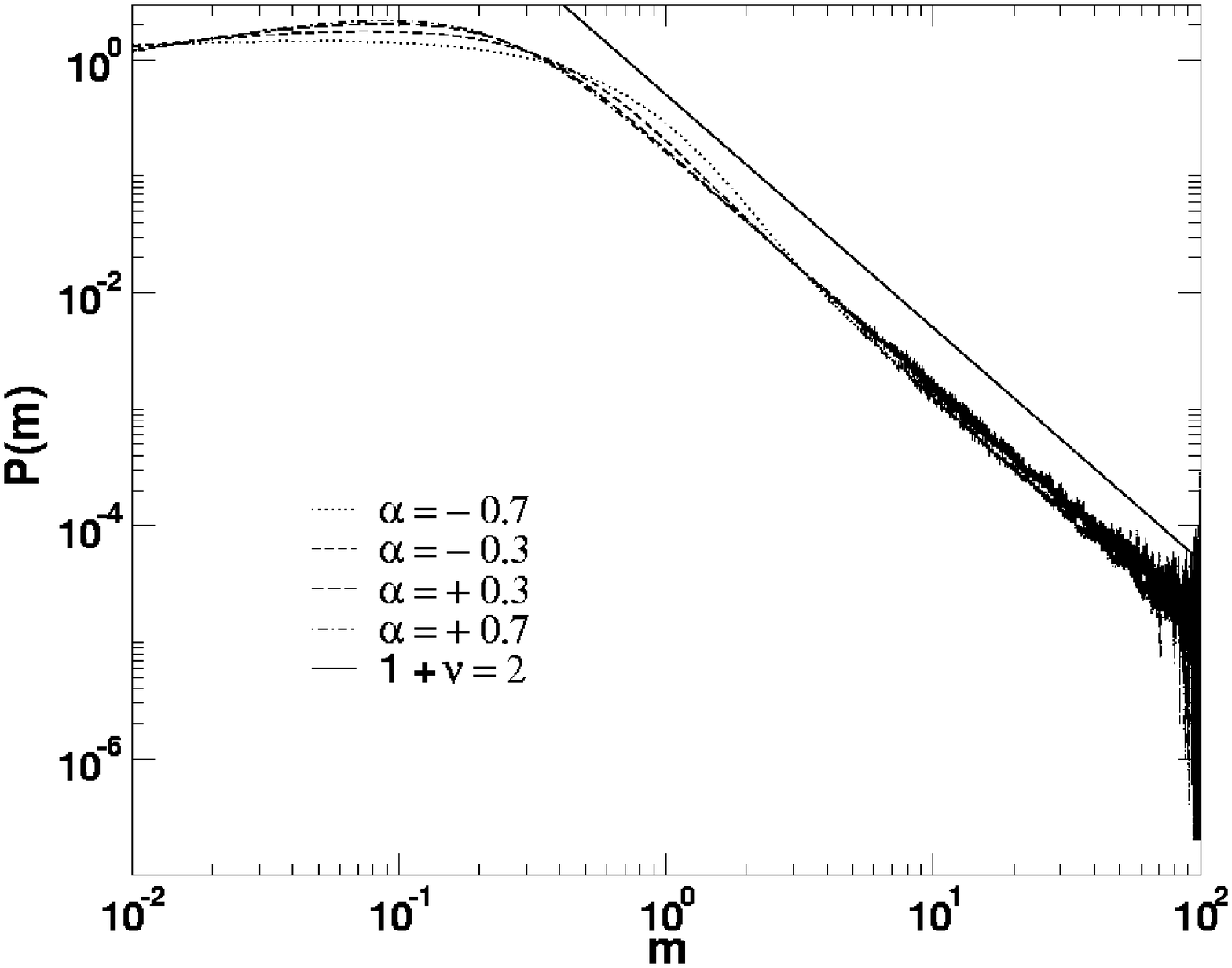}
\caption{Steady state money distribution $P(m)$ in the model with for a 
system of $N=100$ agents with $\lambda$ distributed as 
$\rho(\lambda) \sim \lambda^\alpha$, with different values of $\alpha$.
In all cases, agents play with average money per agent $M/N=1$.}
\label{ccfig:3}       
\end{figure}

In case of uniform distribution of saving propensity $\lambda$
($0 \le \lambda <1$), the individual money distribution $\tilde{P_f}(m)$ for
agents with any particular $\lambda$ value, although differs considerably,
remains non-monotonic: similar to that for fixed $\lambda$ market
with $m_p(\lambda)$ shifting with $\lambda$ (see Fig. 1). Few subtle points may
be noted though: while for fixed $\lambda$ the $m_p(\lambda)$ were all
less than of the order of unity (Fig. 1(a)), for distributed $\lambda$ case
$m_p(\lambda)$ can be considerably larger and can approach to the order of $N$
for large $\lambda$ (see Fig. 1(b)). The other important difference is in the
scaling behavior of $\tilde{P_f}(m)$, as shown in the inset of Fig. 1(b).
In the distributed $\lambda$ ensemble, $\tilde{P_f}(m)$ appears to have a very
simple scaling:
\begin{equation}
\label{scale}
\tilde{P}_f(m) \sim (1-\lambda) \mathcal F (m(1-\lambda)),
\end{equation}

\noindent
for $\lambda \rightarrow 1$, where the scaling function ${\cal F} (x)$ has
non-monotonic variation in $x$. The fixed (same for all agents) $\lambda$
income distribution $P_f(m)$ do not have any such comparative scaling property.
It may be noted that a small difference exists between the
ensembles considered in Fig 1(a) and 1(b): while $\int mP_f(m)dm=M$
(independent of $\lambda$), $\int m \tilde{P_f}(m)dm$ is not a constant and
infact approaches to order of $M$ as $\lambda \rightarrow 1$.
There is also a marked qualitative difference in fluctuations (see Fig. 4):
while for fixed $\lambda$, the fluctuations in time (around the most-probable
value) in the individuals' money $m_i(t)$ gradually decreases with increasing
$\lambda$, for quenched distribution of $\lambda$, the trend gets reversed
(see Fig. 4).
\begin{figure}
\vskip .7cm
\centering
\includegraphics[height=8cm]{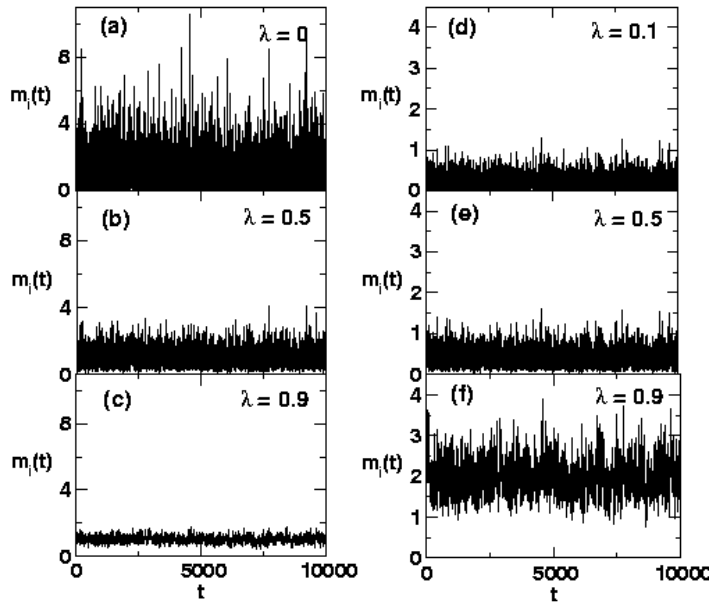}
\caption{
Time variation of the money of the $i$th trader: For fixed $\lambda$
market -- (a), (b), (c); and for agents with specific values of $\lambda$
in the distributed $\lambda$ market -- (d), (e), (f).}
\label{ccfig:4}       
\end{figure}

We now investigate on the range of distribution of the saving propensities in
a certain interval $a<\lambda_i<b$, where, $0<a<b<1$. For uniform distribution
within the range, we observe the
appearance of the same power law in the distribution but for a narrower
region. As may be seen from Fig. 5, as $a \rightarrow b$, the power-law
behavior is seen for values $a$ or $b$ approaching more and more towards unity:
For the same width of the interval $|b-a|$, one gets power-law (with
same $\nu$) when $b \rightarrow 1$. This indicates, for fixed $\lambda$,
$\lambda=0$ corresponds to Gibbs distribution, and one gets Pareto law when
$\lambda$ has got non-zero width of its distribution extending upto
$\lambda = 1$. This of course indicates a crucial
role of these high saving propensity agents: the power law behavior is truely
valid upto the asymptotic limit if $\lambda = 1$ is included. Indeed, had we
assumed $\lambda_0=1$ in (\ref{lam0}),
the Pareto exponent $\nu$ immediately switches over to $\nu=1+\alpha$.
Of course, $\lambda_0 \ne 1$ in (\ref{lam0}) leads to the universality of the
Pareto distribution with $\nu = 1$ (irrespective of $\lambda_0$ and $\alpha$).
Indeed this can be easily rationalised from the scaling behavior (\ref{scale}):
$P(m) \sim \int_0^1 \tilde{P_f}(m)\rho(\lambda)d\lambda$ $\sim$ $m^{-2}$ for
$\rho(\lambda)$ given by (\ref{lam0}) and $m^{-(2+\alpha)}$ if
$\lambda_0=1$ in (\ref{lam0}) (for large $m$ values).
\begin{figure}
\vskip .6cm
\centering
\includegraphics[height=8cm]{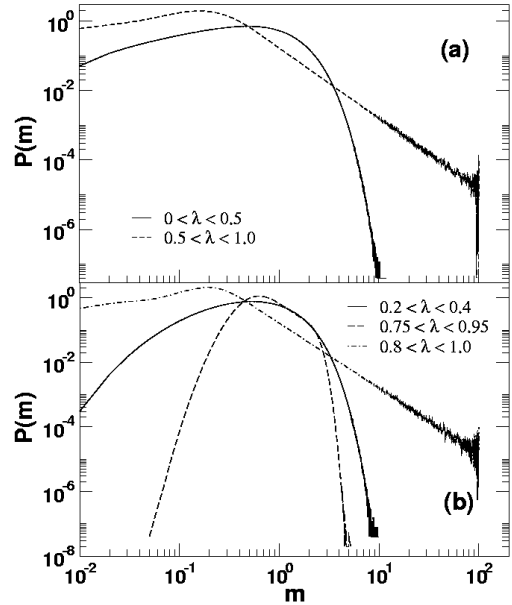}
\caption{Steady state money distribution in cases when the saving propensity 
$\lambda$ is distributed uniformly within a range of values:
(a) width of $\lambda$ distribution is $0.5$, money distribution shows
power law for $0.5 < \lambda < 1.0$;
(a) width of $\lambda$ distribution is $0.2$, money distribution shows
power law for $0.7 < \lambda < 0.9$.
The power law exponent is $\nu \simeq 1$ in all cases. All data shown here are
for $N=100$, $M/N=1$.
}\label{ccfig:5}       
\end{figure}

These model income distributions $P(m)$ compare very well with the wealth 
distributions of various countries: Data suggests Gibbs like distribution 
in the low-income range (more than 90\% of
the population) and Pareto-like in the high-income range \cite{cc:realdatag} 
(less than 10\% of the population) of various countries. 
In fact, we compared one model simulation of the market with saving
propensity of the agents distributed following (\ref{lam0}), with $\lambda_0=0$
and $\alpha=-0.7$ \cite{cc:Chatterjee:2004}. 
The qualitative resemblance of the model income distribution with the real
data for Japan and USA in recent years is quite intriguing. In fact, for
negative $\alpha$ values in (\ref{lam0}), the density of traders with low
saving propensity is higher and since $\lambda=0$ ensemble yields
Gibbs-like income distribution (\ref{gibbs}), we see an initial Gibbs-like
distribution which crosses over to Pareto distribution (\ref{par}) with
$\nu=1.0$
for large $m$ values. The position of the crossover point depends on the
value of $\alpha$. It is important to note that any distribution
of $\lambda$ near $\lambda=1$, of finite width, eventually gives Pareto law
for large $m$ limit. The same kind of crossover behavior (from Gibbs to Pareto)
can also be reproduced in a model market of mixed agents where $\lambda=0$
for a finite fraction of population and $\lambda$ is distributed uniformly over
a finite range near $\lambda=1$ for the rest of the population.

We even considered annealed randomness in the saving propensity $\lambda$:
here $\lambda_i$ for any agent $i$ changes from one value to another within
the range $0 \le \lambda_i < 1$, after each trading. Numerical studies for
this annealed model did not show any power law behavior for $P(m)$; rather
it again becomes exponentially decaying on both sides of a most-probable value.

\section{Dynamics of money exchange}
\noindent
We will now investigate the steady state distribution of money resulting
from the above two equations representing the trading and money dynamics.
We will now solve the dynamics of money distribution in two limits. In one
case, we study the evolution of the mutual money difference among the agents
and look for a self-consistent equation for its steady state distribution.
In the other case, we develop a master equation for the money distribution
function.
\begin{figure}
\centering
\includegraphics[height=6.5cm]{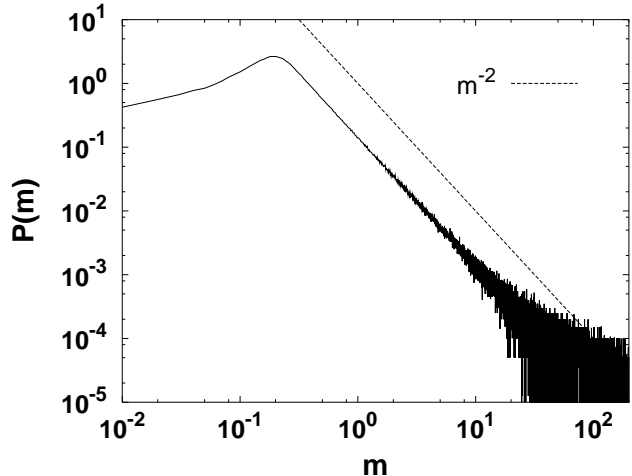}
\caption{
Steady state money distribution $P(m)$ against $m$ in a numerical simulation
of a market with $N=200$, following equations (\ref{mi}) and (\ref{mj})
with $\epsilon_{ij}=1/2$.
The dotted lines correspond to $m^{-(1+\nu)}$; $\nu=1$.}
\label{ccfig:6}       
\end{figure}

\subsection{Distribution of money difference}
\noindent
Clearly in the process as considered above, the total money $(m_i+m_j)$
of the pair of agents $i$ and $j$ remains
constant, while the difference $\Delta m_{ij}$ evolves as
\begin{eqnarray}
\label{dDelmijtt}
(\Delta m_{ij})_{t+1} 
&\equiv& 
(m_i-m_j)_{t+1} \nonumber\\
&=&
\left( \frac{\lambda_i+\lambda_j}{2} \right)(\Delta m_{ij})_t +
\left( \frac{\lambda_i-\lambda_j}{2} \right)(m_i+m_j)_t \nonumber\\
&& +
(2 \epsilon_{ij} -1)[(1-\lambda_i)m_i(t)+(1-\lambda_j)m_j(t)].
\end{eqnarray}

\noindent
Numerically, as shown in Fig. 2, we observe that the steady state money
distribution in the market becomes a power law, following such tradings
when the saving factor $\lambda_i$ of the agents remain constant over time
but varies from agent to agent widely. As shown in the numerical simulation
results for $P(m)$ in Fig. 6, the law, as well as the exponent, remains
unchanged even when $\epsilon_{ij}=1/2$ for every trading.
This can be justified by the earlier numerical observation
\cite{cc:Chakraborti:2000,cc:Chatterjee:2004} for fixed $\lambda$ market
($\lambda_i = \lambda$ for
all $i$) that in the steady state, criticality occurs as $\lambda \to 1$
where of course the dynamics becomes extremely slow. In other words,
after the steady state is realized, the third term in (\ref{dDelmijtt})
becomes unimportant for the critical behavior.
We therefore concentrate on this case, where the above evolution equation for
$\Delta m_{ij}$ can be written in a more simplified form as
\begin{equation}
\label{dDelmt}
(\Delta m_{ij})_{t+1} = \alpha_{ij}(\Delta m_{ij})_t + \beta_{ij}(m_i+m_j)_t,
\end{equation}

\noindent
where $\alpha_{ij}=\frac{1}{2}(\lambda_i+\lambda_j)$ and
$\beta_{ij}=\frac{1}{2}(\lambda_i-\lambda_j)$. As such, $0 \le \alpha < 1$ and
$-\frac{1}{2} < \beta < \frac{1}{2}$.

The steady state probability distribution $D$ for the modulus
$\Delta = |\Delta m|$ of the mutual money difference between any two agents
in the market can be obtained from (\ref{dDelmt}) in the following way
provided $\Delta$ is very much larger than the average money per agent $=M/N$.
This is because, large $\Delta$ can appear from `scattering' involving
$m_i - m_j = \pm \Delta$ and when either $m_i$ or $m_j$ is small. When both
$m_i$ and $m_j$ are large, maintaining a large $\Delta$ between them,
their probability is much smaller and hence their contribution.
Then if, say, $m_i$ is large and $m_j$ is not, the right hand
side of (\ref{dDelmt}) becomes $\sim (\alpha_{ij}+ \beta_{ij})(\Delta_{ij})_t$
and so on. Consequently for large $\Delta$ the distribution $D$ satisfies
\begin{eqnarray}
\label{DDel}
D(\Delta)
&=& \int d \Delta^\prime \; D(\Delta^\prime) \;
\langle 
\delta (\Delta -(\alpha + \beta) \Delta^\prime) +
\delta (\Delta -(\alpha - \beta) \Delta^\prime) 
\rangle \nonumber\\
&=& 
2 \langle 
\left( \frac{1}{\lambda} \right)
\; D
\left( \frac{\Delta}{\lambda} \right)
\rangle,
\end{eqnarray}

\noindent
where we have used the symmetry of the $\beta$ distribution and the relation
$\alpha_{ij} + \beta_{ij}=\lambda_i$, and have suppressed labels $i$, $j$.
Here $\langle \ldots \rangle$ denote average over $\lambda$ distribution
in the market.
Taking now a uniform random distribution of the saving factor $\lambda$,
$\rho(\lambda) = 1$ for $0 \le \lambda < 1$, and assuming
$D(\Delta) \sim \Delta^{-(1+\gamma)}$ for large $\Delta$, we get
\begin{equation}
\label{gammaex}
1=2 \int d \lambda \; \lambda^\gamma = 2 (1+\gamma)^{-1},
\end{equation}

\noindent
giving $\gamma=1$.
No other value fits the above equation. This also
indicates that the money distribution $P(m)$ in the market also follows a
similar power law variation, $P(m) \sim m^{-(1+\nu)}$ and $\nu=\gamma$.
We will now show in a more rigorous way that indeed the only stable solution
corresponds to $\nu=1$, as observed numerically
\cite{cc:Chatterjee:2004,cc:Chatterjee:2003,cc:Chakrabarti:2004}.

\subsection{Master equation and its analysis}
\noindent
We also develop a Boltzmann-like master equation for the time
development of $P(m,t)$, the probability distribution of money in the
market \cite{cc:Chatterjee:2005,cc:Chatterjee:2005a}. We again consider the case
$\epsilon_{ij}=\frac{1}{2}$ in (\ref{mi}) and (\ref{mj}) and rewrite them as
\begin{equation}
\label{Amat1}
\left(
\begin{array}{c}
m_i\\m_j
\end{array}
\right)_{t+1}
= \mathcal{A}
\left(
\begin{array}{c}
m_i\\m_j
\end{array}
\right)_t
{\rm where}\;\;\;
\mathcal{A}=
\left(
\begin{array}{cc}
\mu_i^+ & \mu_j^-\\
\mu_i^- & \mu_j^+
\end{array}
\right);\quad
\mu^\pm = \frac{1}{2} (1 \pm \lambda).
\end{equation}
                                                                                                                             
\noindent
Collecting the contributions from terms scattering in and subtracting
those scattering out, we can
write the master equation for $P(m,t)$ as
\begin{equation}
\label{partial}
\frac{\partial P(m,t)}{\partial t} + P(m,t) =
\langle
\int d m_i \int d m_j \; P(m_i,t)P(m_j,t)\;
\delta(\mu_i^+ m_i + \mu_j^- m_j -m)
\rangle,
\end{equation}
                                                                                                                             
\noindent
which in the steady state gives
\begin{equation}
\label{aftrpartial}
P(m) =
\langle
\int d m_i \int d m_j \;P(m_i)P(m_j)\;
\delta(\mu_i^+  m_i +\mu_j^- m_j -m)
\rangle.
\end{equation}
                                                                                                                             
\noindent
Assuming, $P(m) \sim m^{-(1+\nu)}$ for
$m \rightarrow \infty$, we get \cite{cc:Chatterjee:2005,cc:Chatterjee:2005a}
\begin{equation}
\label{mugama}
1
=
\langle
(\mu^+)^\nu + (\mu^-)^\nu
\rangle
\equiv
\int \int d\mu^+ d\mu^- p(\mu^+) q(\mu^-)
\left[
(\mu^+)^\nu + (\mu^-)^\nu
\right].
\end{equation}
                                                                                                                             
\noindent
Considering now the dominant terms
($\propto x^{-r}$ for $r>0$, or $\propto \ln (1/x)$ for $r=0$) in the
$x \to 0$ limit of the integral
$\int_0^\infty m^{(\nu + r)} P(m) \exp (-mx) dm$, we
get from eqn. (\ref{mugama}), after integrations,
$1=2/(\nu + 1)$, giving finally $\nu=1$ (details in Appendix).
\section{Summary and Discussions}
\label{summary}

\noindent
We have numerically simulated here ideal-gas like models of trading markets,
where each agent is identified with a gas molecule and each trading as an
elastic or money-conserving two-body collision. Unlike in the ideal gas,
we introduce (quenched) saving propensity of the agents, distributed widely
between the agents ($0 \le \lambda < 1$). For quenched random variation of
$\lambda$ among the agents the system remarkably self-organizes to a
critical Pareto distribution (\ref{par}) of money with $\nu \simeq 1.0$
(Fig. 2). The exponent is quite robust: for savings distribution
$\rho(\lambda) \sim |\lambda_0-\lambda|^\alpha$,
$\lambda_0 \ne 1$, one gets the same Pareto law with $\nu = 1$ (independent
of $\lambda_0$ or $\alpha$). 

A master equation for
$P(m,t)$, as in (\ref{partial}), for the original case (eqns. (\ref{mi}) and
(\ref{mj})) was first formulated for fixed $\lambda$ ($\lambda_i$ same for all
$i$), in \cite{cc:Das:2003} and solved numerically. Later, a generalized master
equation for the same, where $\lambda$ is distributed, was formulated
and solved in \cite{cc:Repetowicz:2004} and \cite{cc:Chatterjee:2005}.
We show here that our analytic study 
clearly support the power-law for $P(m)$ with the exponent value
$\nu=1$ universally, as observed numerically earlier
\cite{cc:Chatterjee:2004,cc:Chatterjee:2003,cc:Chakrabarti:2004}.

It may be noted that the trading market model
we have talked about here has got some apparent limitations. The stochastic
nature of trading assumed here in the trading market, through the random
fraction $\epsilon$ in (\ref{eps}), is of course not very straightforward
as agents apparently go for trading with some definite purpose (utility
maximization of both money and commodity). We are however, looking only at
the money transactions between the traders. In this sense, the income
distribution we study here essentially corresponds to `paper money', and not
the `real wealth'. However, even taking money and commodity together, one can
argue (see \cite{cc:Hayes:2002}) for the same stochastic nature of the 
tradings, due to the absence of `just pricing' and the effects of 
bargains in the market.

Apart from the observation that Gibbs (1901) and Pareto (1897)
distributions fall in the same category and can appear naturally in the
century-old and well-established kinetic theory of gas, that this model
study indicates the appearance of self-organized criticality in
the simplest (gas) model so far, when the stability effect of savings
incorporated, is remarkable.


\section{Acknowledgments}
\noindent
We are grateful to A. Chakraborti, S. Pradhan, S. S. Manna and 
R. B. Stinchcombe for collaborations at various stages of our study.

\appendix
\section{Alternative solution of the steady state master equation (\ref{aftrpartial})}

\noindent
Let $S_r (x)= \int_0^\infty dm P(m) m^{\nu+r} \exp(-mx)$; $r \ge 0, x > 0$.
If $P(m) = A/m^{1+\nu}$, then
\begin{eqnarray}
\label{app1}
S_r (x)
&=& 
A \int_0^\infty dm \; m^{r-1} \exp(-mx) \nonumber\\
&\sim&
A \; \frac{x^{-r}}{r} \;\;\;\;\;\;\;\;\; {\rm if} \; \; r > 0 \nonumber\\
&\sim&
A \ln \left( \frac{1}{x} \right)\; \;  \; \; {\rm if} \; \; r = 0.
\end{eqnarray}

\noindent
From eqn. (\ref{aftrpartial}), we can write
\begin{eqnarray}
\label{app2}
&& S_r(x) = \nonumber\\
&&
\langle
\int_0^\infty dm_i \int_0^\infty dm_j \; P(m_i) P(m_j)
(m_i \mu_i^+ + m_j \mu_j^-)^{\nu+r}
\exp[-(m_i \mu_i^+ + m_j \mu_j^-)x]
\rangle \nonumber\\
&\simeq&
\int_0^\infty dm_i \; A m_i^{r-1}
\langle
\exp(-m_i \mu_i^+ x) \left(\mu_i^+ \right)^{\nu + r}
\rangle
\left[
\int_0^\infty dm_j \; P(m_j)
\langle
\exp(-m_j \mu_j^- x) 
\rangle
\right] \nonumber\\
&+&
\int_0^\infty dm_j \; A m_j^{r-1}
\langle
\exp(-m_j \mu_j^- x) \left(\mu_j^- \right)^{\nu + r}
\rangle
\left[
\int_0^\infty dm_i \; P(m_i)
\langle
\exp(-m_i \mu_i^+ x) 
\rangle
\right] \nonumber\\
&&
\end{eqnarray}

\noindent
or,
\begin{eqnarray}
\label{app3}
S_r(x)
&=&
\int_{\frac{1}{2}}^1 d\mu_i^+ \; p(\mu_i^+)
\left( 
\int_0^\infty d m_i \; A m_i^{r-1} \exp(-m_i \mu_i^+ x)
\right)
\left( 
\mu_i^+
\right)^{\nu+r} \nonumber \\
&& +
\int_0^{\frac{1}{2}} d\mu_j^- \; q(\mu_j^-)
\left( 
\int_0^\infty d m_j \; A m_j^{r-1} \exp(-m_j \mu_j^- x)
\right)
\left( 
\mu_j^-
\right)^{\nu+r},
\end{eqnarray}

\noindent
since for small $x$, the terms in the square brackets in (\ref{app2})
approach unity. We can therefore rewrite (\ref{app3}) as
\begin{equation}
\label{app4}
S_r (x) = 2 \left[
\int_{\frac{1}{2}}^1 d \mu^+ (\mu^+)^{\nu + r} S_r (x \mu^+)
+
\int_0^{\frac{1}{2}} d \mu^- (\mu^-)^{\nu + r} S_r (x \mu^-)
\right].
\end{equation}

Using now the forms of $S_r(x)$ as in (\ref{app1}), and collecting terms
of order $x^{-r}$ (for $r>0$) or of order $\ln (1/x)$ (for $r=0$)
from both sides of (\ref{app4}), we get (\ref{mugama}).


\end{document}